\documentclass[twocolumn,english,transaction]{IEEEtran}
\usepackage[T1]{fontenc}
\usepackage[utf8]{luainputenc}
\usepackage{float}
\usepackage{amsmath}
\usepackage{amssymb}
\usepackage{stackrel}
\usepackage{graphicx}
\usepackage{xcolor}

\makeatletter

\providecommand{\tabularnewline}{\\}
\floatstyle{ruled}
\newfloat{algorithm}{tbp}{loa}
\providecommand{\algorithmname}{Algorithm}
\floatname{algorithm}{\protect\algorithmname}

\usepackage{amsthm}
\usepackage{setspace}
\usepackage{url}
\usepackage{caption}
\usepackage{subfigure}
\usepackage{stfloats}

\providecommand{\tabularnewline}{\\}
\floatstyle{ruled}
\newfloat{algorithm}{tbp}{loa}
\providecommand{\algorithmname}{Algorithm}
\floatname{algorithm}{\protect\algorithmname}

\IEEEoverridecommandlockouts
\usepackage{ifpdf}
\usepackage{cite}
\ifCLASSOPTIONcompsoc
\else
\fi

\usepackage{babel}
\renewcommand{\figurename}{Fig.}

\makeatother

\usepackage{babel}
\begin{document}
\title{Downlink Power Minimization in Intelligent Reconfigurable Surface-Aided Security Classification Wireless Communications System}
\author{Jintao Xing, \IEEEmembership{Student Member,~IEEE}, Tiejun~Lv, \IEEEmembership{Senior Member,~IEEE},
Yashuai Cao, \IEEEmembership{Member,~IEEE},\\Jie Zeng, \IEEEmembership{Senior Member,~IEEE}, and Pingmu Huang
\thanks{Manuscript received September 1, 2021; revised March 28, 2022; accepted June 8, 2022. This work was supported by the National Natural Science Foundation of China (No. 62001264), and the Natural Science Foundation of Beijing (No. L192025). \emph{(Corresponding author: Tiejun Lv and Jie Zeng)}.}
\thanks{J. Xing, T. Lv and Y. Cao are with the School of Information and Communication Engineering,
Beijing University of Posts and Telecommunications (BUPT), Beijing
100876, China (e-mail: \{jintaoxing, lvtiejun, yashcao\}@bupt.edu.cn).}
\thanks{J. Zeng is with the School of Cyberspace Science and Technology, Beijing Institute of Technology, Beijing 100081, China (e-mail: zengjie@bit.edu.cn).}
\thanks{P. Huang is with the School of Artificial Intelligence, Beijing University of Posts and Telecommunications (BUPT), Beijing
100876, China (e-mail: pmhuang@bupt.edu.cn).}
}

\maketitle
\begin{abstract}
User privacy protection is considered a critical issue in wireless networks, which drives the demand for various secure information interaction techniques. In this paper, we introduce an intelligent reflecting surface (IRS)-aided security classification wireless communication system, which reduces the transmit power of the base station (BS) by classifying users with different security requirements. Specifically, we divide the users into confidential subscribers with secure communication requirements and general communication users with simple communication requirements. During the communication period, we guarantee the secure rate of the confidential subscribers while ensuring the service quality of the general communication users, thereby reducing the transmit power of the BS. To realize such a secure and green information transmission, the BS implements a beamforming design on the transmitted signal superimposed with artificial noise (AN) and then broadcasts it to users with the assistance of the IRS's reflection. We develop an alternating optimization framework to minimize the BS downlink power with respect to the active beamformers of the BS, the AN vector at the BS, and the reflection phase shifts of the IRS. A successive convex approximation (SCA) method is proposed so that the nonconvex beamforming problems can be converted to tractable convex forms. The simulation results demonstrate that the proposed algorithm is convergent and can reduce the transmit power by 20\% compared to the best benchmark scheme.
\end{abstract}


\section{Introduction}

\global\long\def\figurename{Fig.}%

Information security and user privacy have become increasingly important in fifth-generation (5G) mobile networks due to the explosive growth of wireless terminal devices \cite{8792139}. The inherent exposure of the wireless communication medium inevitably incurs information leakage, which exacerbates wireless network security issues such as data uploading, caching, and secure wireless computing for private data (SWCPD) \cite{9023459}. Consequently, methods for protecting the secure delivery of information during transmission have recently drawn considerable attention. Key encryption technologies have been proposed to maintain wireless security. However, due to improvements in modern computer technology, the traditional security solutions that rely heavily on computing power have become increasingly insecure; it has become easier for eavesdroppers to decipher the security information using supercomputers. Therefore, physical layer security (PLS) was developed as a mainstream wireless technology. The main idea underlying PLS involves utilizing transmission channel characteristics from the physical level. To capitalize on the multiplexing gain brought by large-scale multiple-input multiple-output (MIMO) antennas, various approaches have been developed that employ beamforming at the base station (BS) and introduce additional artificial noise (AN) to enhance the PLS \cite{8962162,9028250}. In the context of large-scale MIMO systems, the beamforming gain attained by high-directional antennas can increase the desired power for confidential subscribers; hence, the effective information received by the eavesdroppers can be greatly reduced.

However, installing massive antennas and radio frequency (RF) chains in large-scale MIMO systems results in excessive hardware costs and signal processing complexity \cite{benzaghta2021massive}. On the one hand, that outcome violates the critical design metric for green communication systems \cite{ozyurt2021intracell}; on the other hand, the high energy consumption involved in active MIMO restricts the practical deployment of MIMO secure wireless communications in 5G and beyond. During the transmission period, using relays to forward signals will lead to large power consumption overhead. Moreover, employing a malicious relay poses a high risk of information leakage \cite{wu2021uav}. These problems urgently need new technology. Recently, the development of electronic circuits and metamaterials technology has allowed the intelligent reflecting surface (IRS) to be exploited in existing wireless networks. The IRS is a plane composed of a large number of low-cost passive reflective elements. To precisely control the reflection direction of the incident signals, each element is connected to a smart controller that can tune its amplitude and phase \cite{8910627}. As a passive forwarding node, the deployment of the IRS offers promise to meet low-cost spectrum efficiency goals \cite{9122596,9198125,9206080}. An IRS can easily be placed on a wall, ceiling, or other area thanks to its light weight and adjustability. The introduction of the IRS offers increased flexibility and extends the possibilities of current wireless systems, which differ from the existing active relays.

Motivated by the aforementioned considerations, this paper considers an IRS-aided secure transmission system in which the passive beamforming technique of the IRS is integrated with the active beamforming and AN interference capabilities of the BS. This paper focuses primarily on how to benefit system performance by adding the new degree of freedom presented by the IRS.

\subsection{Related Works}

Using 5G and beyond, the popular Internet of Things (IoT) and machine learning
applications have intensified the need for secure information transmission
and user privacy protections \cite{8543573,9070153,8758230,8437135}.
Requiring only knowledge of the fundamental radio propagation characteristics,
PLS techniques can safeguard confidential information based on
advanced information theoretic methods \cite{8335290}. Wyner $\emph{et al.}$
\cite{6772207} proposed the concept of a wireless wire-tap model
in 1975, revealing that confidential broadcasting can be
achieved through precoding methods that utilize the physical layer
differences between the legitimate users and the eavesdroppers' channels
at the transmitter. When the channel quality of a legitimate user
is stronger than that of an eavesdropper, the coding methods can be
designed to achieve a positive secrecy rate. Later, S. Zhao $\emph{et al.}$
\cite{9115675} extended the secure beamforming scheme to a full-duplex
MIMO untrusted relay system that maximizes the secrecy sum rate (SSR) of the
system and concluded that the untrusted relay
needs to be treated as a pure eavesdropper under low signal-to-noise
ratio (SNR) conditions to achieve the satisfactory system gain. P.
Huang $\emph{et al.}$ \cite{8680631} investigated the secure beamforming
design in downlink IoT systems. To maximize the SSR, two optimization
schemes were proposed to address scenarios involving both passive and active
eavesdroppers. Superimposing the AN on the transmit signal
has been considered an efficient method for reducing the SNR of passive
eavesdroppers. In \cite{8533374}, the authors explored a novel hierarchical
PLS model and proposed an AN-assisted beamforming scheme to
solve the non-convex SSR maximization problem. S. Hong $\emph{et al.}$ \cite{9201173} proposed an efficient algorithm and derived the optimal TPC matrix, AN covariance matrix by using the Lagrange multiplier method, and the optimal phase shifts by an efficient MM algorithm in secure MIMO wireless communication system. Apart from the conventional orthogonal multiple
access (OMA) mode assumed in the above mentioned studies, S. Huang
$\emph{et al.}$ \cite{9169836} proposed a minimum-angle difference based user pairing scheme in mmWave networks by invoking the non-orthogonal
multiple access (NOMA) technique. Their scheme minimized the secrecy
outage probability of user pairs, and enhanced the confidentiality
performance by designing a hybrid beamforming scheme. In terms of passive defense, H. Zhang $\emph{et al.}$ \cite{9072320} devised
a different strategy by actively listening to two suspicious communication
links to improve the SSR. Additionally, network coding schemes have
been applied to protect information transmission. In \cite{8417403},
the authors proposed generating a secure network coding from multipath
channels by constructing a secure network coding system. A secure
information transmission problem was studied and solved by utilizing
an adaptive quantification algorithm to prevent wiretap attacks.

However, when relays are adopted for auxiliary communications, whether the relays are malicious is generally unknown. When malicious
relays are present, adopting a conventional PLS scheme would degrade the
system performance. Unlike relays, the IRS opens up substantial additional possibilities
for secure transmission due to its passive and portable characteristics
\cite{9149709,9120206} and has been extensively studied for assisting wireless transmission \cite{9475159,9306896}. An IRS-assisted secure wireless communication
system was first considered in \cite{8723525}, which maximized the secrecy
rate of the legitimate communication link by jointly
designing the transmit beamformer at the BS and the passive beamformer
of the IRS. In \cite{9146170}, the authors considered maximizing
the achievable security sum rate in IRS-assisted cognitive radio systems.
To maximize system confidentiality, J. Chen $\emph{et al.}$ \cite{8742603}
formulated a minimum secrecy rate maximization problem under various
practical constraints on the reflecting coefficients, which captures
the nature of phase shifts and amplitudes of the IRS. In \cite{9159923},
the authors considered an IRS-assisted Gaussian MIMO wire tap channel
scenario and found an interesting trade-off between the quality of service
(QoS) at the receiver and the secrecy rate. In \cite{9086467},
the authors considered a more challenging scenario where the eavesdropper
is configured with multiple antennas. Specifically, they proposed
an alternating optimization algorithm that can be extended to the
IRS reflection pattern with discrete phase shifts.

The explosive growth of information must be transmitted through wireless channels, which will inevitably contain a lot of information with high-security performance requirements. It is unrealistic that all communication information is roughly transmitted securely and will cause link redundancy and a considerable waste of resources. Different wireless terminal devices have different requirements for confidentiality in a practical wireless communication system. It is particularly critical to ensure information that involves security and privacy, such as financial transactions or encrypted calls. For instance, if the wireless terminal device requests some irrelevant information, such as entertainment media resources or current public news, and the information does not need to be encrypted. In a wireless application protocol (WAP), users with different security requirements write their requests into data packets and send them to the BS \cite{erlandson2000wap}. The BS divides the user into confidential subscribers or general communication users according to its security request type. The transmission of highly confidential information for confidential subscribers often requires a higher service priority, so operators can use different security transmission schemes and implement different pricing strategies to achieve economic gain. The security classification scheme can not only effectively reduce the redundancy of the secure encryption during information transmission but also increase the operator's economic benefits.

\subsection{Contributions}

Most studies assume that the same secure transmission policies are
applied to all user devices in a cellular system. However, this secure
communication mechanism will lead to redundant operations for some communication users without security
requirements, resulting in additional energy expenditure. In contrast to existing works,
we consider the different security requirements of users
in the network; thus, unnecessary energy consumption may be saved and green
communication realized by performing a security classification
for all users. Specifically, our scheme divides all communication users into two
categories based on their security communications
requirements: confidential subscribers and general communication
users. Confidential subscribers consist of users that need to
communicate securely with the BS due to their own communication confidentiality
requirements, or to the existence of a terminal that may potentially attempt to steal
the transmission signals. General communication users refer to
those users who do not require confidential transmission; only their
QoS constraints need to be met during the communication period. In
the considered IRS-aided security classification wireless communication
system, we expect to achieve secure and green communication goals
by developing an alternating optimization framework to improve the
security performance of the legitimate communication users. We aim
to minimize the downlink transmit power of the BS while satisfying
users' different security rate requirements by optimizing the
transmit beamforming vectors at the BS, the injected AN vector at
the BS, and the reflection phase shifts of the IRS. The
key contributions of this paper are summarized as follows.
\begin{itemize}
\item We consider a new security classification scheme in an IRS-aided wireless communication system. By classifying users into confidential subscribers with secure communication requirements and general communication users with simple communication requirements, we formulate joint active beamforming, AN injection and IRS passive beamforming problems to improve the transmit power utilization.
\item The formulated active power minimization problem is difficult to solve since it is a multivariate optimization problem with multiple nonconvex constraints. We propose an alternating optimization framework to divide the original problem into a series of subproblems. Then, we derive the first-order Taylor approximation expressions of the nontrivial decoupled problems to make them tractable.
\item We reformulate the passive beamforming subproblem into rank-one constrained linear matrix inequality (LMI) problems to address the unit modulus constraints; hence, the optimal phase shift vector can be recovered by invoking the eigen value decomposition (EVD) with Gaussian randomization. In addition, we analyze the convergence of the proposed algorithm. The simulation results reveal insights about the optimal deployments of the IRS and communication users and verify the effectiveness of the proposed algorithm.
\end{itemize}
\ \
The remainder of this paper is organized as follows. In Section II, we present
the IRS-aided security classification wireless communication system.
In Section III, we analyze and reformulate the optimization problem,
followed by the proposed power minimization algorithm. The numerical
results and discussion are provided in Section IV. Finally, Section V
concludes this paper.

\emph{Notations}: Boldface lower and uppercase letters denote
vectors and matrices, respectively. The Hermitian transpose, Frobenius
norm, and the trace of the matrix $\mathbf{A}$ are denoted as $\mathbf{A}^{H}$,
$\left\Vert \mathbf{A}\right\Vert $, and $\mathrm{Tr}(\mathbf{A})$,
respectively. $|\mathbf{A}|$ stands for the determinant of $\mathbf{A}$.
$\mathrm{Diag}\left(\mathbf{a}\right)$ creates a diagonal matrix
with the entries of vector $\mathbf{a}$ on its diagonal. $\left[\cdot\right]^{+}\triangleq\mathrm{max}\left\{ 0,\,\cdot\right\} $.
$\mathcal{CN}\left(\mu,\sigma^{2}\right)$ denotes a circularly
symmetric complex Gaussian distribution with mean $\mu$ and variance
$\sigma^{2}$.

\section{System Model}

\begin{figure}
\centering{}\includegraphics[scale=0.4]{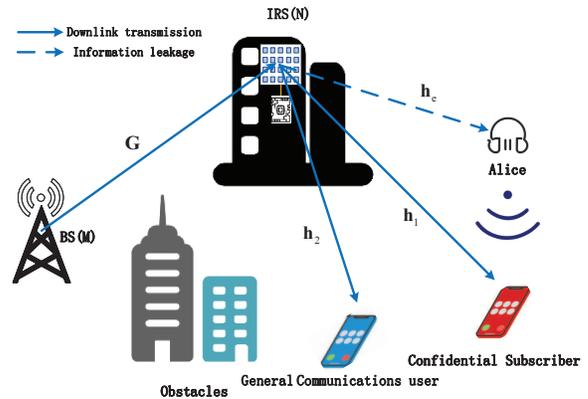}\caption{An illustration of the considered scenario.}
\end{figure}

As shown in Fig. 1, we consider an IRS-aided security classification wireless communication system. The system consists of a BS, an IRS, a confidential subscriber, a general communication user and an eavesdropper named Alice. The confidential subscriber needs to request confidential information such as financial transactions or encrypted calls and has requirements carved into data packets to transmit to the BS. The BS classifies this user as a confidential subscriber by identifying the data packets according to the WAP. Similarly, a user with nonconfidential information such as entertainment videos or current affairs news is classified as a general communication user. The eavesdropper named Alice tries to eavesdrop on vital information for the confidential subscriber. We assume that there are obstacles
(such as tall buildings, hillsides, etc.) between the users and the
BS. Therefore, the BS cannot transmit information directly to
users; it can communicate with users only through the reflection
links of the IRS. The eavesdropper is assumed to be passive,
which means that it can receive only the broadcast information due to the open wireless transmission medium and cannot actively send interference signals or misroute the transmitted
data. The BS is equipped with $M$ transmit antennas. The IRS unit integrates a large number of low-cost meta-atoms, with $N_{a}$ elements horizontally and $N_{b}$ elements vertically arranged. The total number of passive elements of the IRS is given by $N=N_{a}\times N_{b}$. In the considered system, we assume that the confidential subscriber, the
general communication user, and Alice are each equipped with a single antenna.
All the nodes work in half-duplex mode and all the channels involved
in this paper are assumed to be quasi-static flat-fading. The IRS-aided
PLS works assuming that perfect CSI is available and widely considered
\cite{yu2021irs,wang2020energy}. The CSI between the IRS and devices
can be acquired by resorting to classic and high-precision channel
estimation methods such as alternating least squares (ALS) and vector
approximate message passing (VAMP) methods \cite{wei2021channel,hu2021two}.

Let $\mathbf{G}\in\mathbb{C}^{N\times M}$ denote the baseband equivalent
channels between the BS and the IRS, and let $\mathbf{h}_{1}$, $\mathbf{h}_{2}$
and $\mathbf{h}_{e}$ denote that between the IRS and the confidential
subscriber, the IRS and the general communication user, and the IRS and
Alice, respectively. We define the diagonal reflection phase shift matrix
as $\mathbf{\boldsymbol{\Phi}}=\mathrm{diag}\left(\mathbf{v}\right)$,
where $\mathbf{v}=\left[v_{1},v_{2},\cdots,v_{N}\right]^{T}$, $v_{n}=e^{j\theta_{n}}$
interprets the unit modulus reflection coefficient corresponding to
the $n$-th IRS element, and $\theta_{n}\in[0,2\pi)$ denotes the
reflection coefficient of the $\ensuremath{n}$-th element. The symbol
notations and their explanations are listed in Table I.

\begin{table}
\caption{PARAMETERS AND THEIR MEANINGS}

\centering{}%
\begin{tabular}{cc}
\hline
$\mathbf{w}_{1}$ & Beamforming for Confidential Subscriber\tabularnewline
$\mathbf{w}_{2}$ & Beamforming for Communication User\tabularnewline
$\mathbf{z}$ & Artificial Noise Vector\tabularnewline
$s$ & Transmit Symbol for Users\tabularnewline
$P_{t}$ & Total Transmit Power\tabularnewline
$\mathbf{G}$ & CSI of BS to IRS\tabularnewline
$\mathbf{h}_{1}$ & CSI of BS to Confidential Subscriber\tabularnewline
$\mathbf{h}_{2}$ & CSI of BS to Communication User\tabularnewline
$\mathbf{h}_{e}$ & CSI of IRS to Alice\tabularnewline
$\mathbf{\boldsymbol{\Phi}}$  & Reflection Angle of IRS\tabularnewline
$\sigma^{2}$ & Noise Power\tabularnewline
\hline
\end{tabular}
\end{table}

The BS broadcasts a superposition of the source data and the AN symbols
to the users through the active beamforming design. The transmit signal
at the BS is expressed as

\begin{align}
\mathbf{x} & =\mathbf{w}_{1}s_{1}+\mathbf{w}_{2}s_{2}+\mathbf{z},\nonumber \\
 & =\left[\mathbf{w}_{1},\mathbf{w}_{2},\mathbf{z}\right]\cdot\left[\begin{array}{c}
s_{1}\\
s_{2}\\
1
\end{array}\right],
\end{align}
where $s_{1}$ and $s_{2}$ indicate the information-bearing signal
to the confidential subscriber and general communication user,
respectively. The corresponding beamforming vectors are denoted as
$\mathbf{w}_{i}\in\mathbb{C}^{M\times1}$, and $i\in\{1,2\}$. $\mathbf{w}_{1}$
denotes the beamforming vector corresponding to the confidential subscriber
and is also termed as the ``confidential beamformer''. $\mathbf{w}_{2}$
denotes the beamforming vector corresponding to the general communication
user and is also called the ``general beamformer''. The considered
signal $s_{i}$ is the unit power, i.e., $\mathbb{E}[|s_{i}|^{2}]=1$.
The AN $\mathbf{z}\in\mathbb{C}^{M\times1}$ is generated from the
distribution of $\mathrm{\mathcal{CN}\left(0,\mathbf{Z}\right)}$
with $\mathbf{Z}\succeq0$ being the covariance matrix of the AN vector
\cite{8680631}.

The signals received by the confidential subscriber and the general
communication user are denoted by
\begin{equation}
y_{c} =\mathbf{h}_{1}^{H}\boldsymbol{\Phi}\mathbf{G}\left(\mathbf{w}_{1}s_{1}+\mathbf{w}_{2}s_{2}+\mathbf{z}\right)+n_{1},
\end{equation}
and
\begin{equation}
y_{g} =\mathbf{h}_{2}^{H}\mathbf{\boldsymbol{\Phi}}\mathbf{G}\left(\mathbf{w}_{1}s_{1}+\mathbf{w}_{2}s_{2}+\mathbf{z}\right)+n_{2},
\end{equation}
respectively, where $n_{k}\sim\mathrm{\mathcal{CN}}\left(0,\sigma_{k}^{2}\right)$
represents additive white Gaussian noise (AWGN) and $\sigma_{k}^{2}$
is the corresponding noise variance.

Because Alice is primarily interested in intercepting confidential information from the
subscriber, the signal received by Alice is expressed as\footnote{Generally, eavesdroppers often act as regular communication
users in cellular networks and perform eavesdropping attacks only during a certain task period.}

\begin{equation}
y_{e}=\mathbf{h}_{e}^{H}\boldsymbol{\Phi}\mathbf{G}\left(\mathbf{w}_{1}s_{1}+\mathbf{w}_{2}s_{2}+\mathbf{z}\right)+n_{e}.
\end{equation}

Accordingly, the data rates of the confidential subscriber, the general
communication user and Alice are given by\footnote{In the IRS-aided communication system, the BS and IRS are usually
installed on a building tower or floor that is relatively high above
the ground; thus, it is unrealistic for eavesdroppers to deploy interception
equipment at high altitudes. Therefore, Alice receives only the
signals reflected by the IRS, not the modulated signal sent over direct BS-user links.}

\begin{align}
R_{c} & =\log_{2}\left(1+\frac{|\mathbf{h}_{1}^{H}\boldsymbol{\Phi}\mathbf{G}\mathbf{w}_{1}|^{2}}{|\mathbf{h}_{1}^{H}\boldsymbol{\Phi}\mathbf{G}\left(\mathbf{w}_{2}+\mathbf{z}\right)|^{2}+\sigma_{1}^{2}}\right),\\
R_{g} & =\log_{2}\left(1+\frac{|\mathbf{h}_{2}^{H}\mathbf{\boldsymbol{\Phi}}\mathbf{G}\mathbf{w}_{2}|^{2}}{|\mathbf{h}_{2}^{H}\mathbf{\boldsymbol{\Phi}}\mathbf{G}\left(\mathbf{w}_{1}+\mathbf{z}\right)|^{2}+\sigma_{2}^{2}}\right),
\end{align}
and
\begin{equation}
R_{e} =\log_{2}\left(1+\frac{|\mathbf{h}_{e}^{H}\boldsymbol{\Phi}\mathbf{G}\mathbf{w}_{1}|^{2}}{|\mathbf{h}_{e}^{H}\boldsymbol{\Phi}\mathbf{G}\left(\mathbf{w}_{2}+\mathbf{z}\right)|^{2}+\sigma_{e}^{2}}\right),
\end{equation}
respectively.

According to Wyner's theory \cite{6772207}, the SSR of the confidential
subscriber, denoted by $R_{sec}$, can be formulated as (\ref{eq:Security-Rate of confidential user}),
shown at the top of the next page, where $\left[x\right]{}^{+}=\max\left\{ x,0\right\} $.

\begin{figure*}
\begin{align}
R_{sec} & =\left[R_{c}-R_{e}\right]{}^{+}\nonumber \\
 & \triangleq\left[\log_{2}\left(1+\frac{|\mathbf{h}_{1}^{H}\boldsymbol{\Phi}\mathbf{G}\mathbf{w}_{1}|^{2}}{|\mathbf{h}_{1}^{H}\boldsymbol{\Phi}\mathbf{G}\left(\mathbf{w}_{2}+\mathbf{z}\right)|^{2}+\sigma_{1}^{2}}\right)-\log_{2}\left(1+\frac{|\mathbf{h}_{e}^{H}\boldsymbol{\Phi}\mathbf{G}\mathbf{w}_{1}|^{2}}{|\mathbf{h}_{e}^{H}\boldsymbol{\Phi}\mathbf{G}\left(\mathbf{w}_{2}+\mathbf{z}\right)|^{2}+\sigma_{e}^{2}}\right)\right]^{+},\label{eq:Security-Rate of confidential user}
\end{align}
\hrulefill
\end{figure*}

Our goal in this paper is to meet both the communication requirements
of the confidential subscriber and the basic QoS constraints of the general
communication user who lacks security requirements. Mathematically,
the problem is formulated as follows:
\begin{align}
\mathbf{P1:}\qquad & \underset{\mathbf{w}_{1},\mathbf{w}_{2},\mathbf{z},\boldsymbol{\Phi}}{\min}\;\;P_{t}\nonumber \\
s.t.\qquad & \mathrm{C1}:R_{sec}\geq\delta_{1},\nonumber \\
 & \mathrm{C2}:R_{g}\geq\delta_{2},\\
 & \mathrm{C3}:\mathbf{Z}\succeq0,\nonumber \\
 & \mathrm{C4}:\mathbf{\boldsymbol{\Phi}}=\mathrm{diag}\left(\mathbf{v}\right),\vert v_{n}\vert^{2}=1,\nonumber
\end{align}
where $P_{t}=\Vert\mathbf{w}_{1}\Vert^{2}+\Vert\mathbf{w}_{2}\Vert^{2}+\Vert\mathbf{z}\Vert^{2}$
denotes the total power required at the BS. Constraint
C1 indicates that the security rate of the confidential subscriber
should satisfy a minimum security requirement $\delta_{1}$. Constraint
C2 is imposed to satisfy the general communication user's minimum
communication transmission QoS requirement $\delta_{2}$ and implies
that the confidentiality performance does not need to be considered for general communication users. Constraint C3 specifies that the covariance
matrix of the AN is a Hermitian matrix and positive semidefinite.
Constraint C4 specifies the reflection phase shifts of the IRS as
a diagonal matrix with $N$ unit modulus entries.

\section{Algorithm Design for Secure IRS-aided Wireless Communication}

Problem $\mathbf{P1}$ cannot be directly solved using the traditional
convex optimization tools because it is difficult to optimize multiple
optimization variables simultaneously due to the coupling variables
and the non-convexity of the constraints. To decouple these optimization
variables, we develop an alternating optimization algorithm to optimize
the active confidential beamformer $\mathbf{w}_{1}$ and general beamformer
$\mathbf{w}_{2}$ at the BS, the AN, and the IRS coefficient matrix
in an iterative manner, which can achieve the suboptimal solution of the
original problem. Based on the alternating optimization framework,
the original problem can be decomposed into three subproblems, i.e.,
the optimization of the confidential beamformer, the joint optimization
of the general beamformer and the design of the injected AN vector,
and the optimization of the IRS phase shifts. By solving these subproblems
in an alternating iterative manner, the overall transmit power can
be minimized to realize secure and green communications.

\subsection{Optimization of the Confidential Beamformer $\mathbf{w}_{1}$}

By taking the structural characteristics of the expression
constraints C1 and C2 into account, we observe that $\mathbf{w}_{1}$ and $\mathbf{w}_{2}$
are multiplicatively coupled in the optimization problem $\mathbf{P1}$.
Therefore, we optimize $\mathbf{w}_{1}$ and $\mathbf{w}_{2}$ separately.

Given the general beamformer $\mathbf{w}_{2}$, the AN vector $\mathbf{z}$
and the reflection coefficients $\boldsymbol{\Phi}$ of the IRS, only
constraints C1 and C2 are involved in the optimization of $\mathbf{w}_{1}$;
consequently, constraints C3 and C4 can be omitted. Note that C1 and C2
are nonconvex with respect to $\mathbf{w}_{1}$ when fixing the variables
$\mathbf{w}_{2},\boldsymbol{\Phi}$ and $\mathbf{z}$. To deal with the
nonconvexity, C1 can be transformed as follows:

\begin{align} \mathrm{C1a}:&R_{sec}=\log_{2}\left(1+\frac{|\mathbf{h}_{1}^{H}\boldsymbol{\Phi}\mathbf{G}\mathbf{w}_{1}|^{2}}{|\mathbf{h}_{1}^{H}\boldsymbol{\Phi}\mathbf{G}\left(\mathbf{w}_{2}+\mathbf{z}\right)|^{2}+\sigma_{1}^{2}}\right)\nonumber
 \\
 & 	
 -\log_{2}\left(1+\frac{|\mathbf{h}_{e}^{H}\boldsymbol{\Phi}\mathbf{G}\mathbf{w}_{1}|^{2}}{|\mathbf{h}_{e}^{H}\boldsymbol{\Phi}\mathbf{G}\left(\mathbf{w}_{2}+\mathbf{z}\right)|^{2}+\sigma_{e}^{2}}\right)\geq\delta_{1}\nonumber
 \\
 &
 = \frac{|\mathbf{h}_{1}^{H}\boldsymbol{\Phi}\mathbf{G}\mathbf{w}_{1}|^{2}+|\mathbf{h}_{1}^{H}\boldsymbol{\Phi}\mathbf{G}\left(\mathbf{w}_{2}+\mathbf{z}\right)|^{2}+\sigma_{1}^{2}}{|\mathbf{h}_{e}^{H}\boldsymbol{\Phi}\mathbf{G}\mathbf{w}_{1}|^{2}+|\mathbf{h}_{e}^{H}\boldsymbol{\Phi}\mathbf{G}\left(\mathbf{w}_{2}+\mathbf{z}\right)|^{2}+\sigma_{e}^{2}}\nonumber
\\
 & 	\times\frac{|\mathbf{h}_{e}^{H}\boldsymbol{\Phi}\mathbf{G}\left(\mathbf{w}_{2}+\mathbf{z}\right)|^{2}+\sigma_{e}^{2}}{|\mathbf{h}_{1}^{H}\boldsymbol{\Phi}\mathbf{G}\left(\mathbf{w}_{2}+\mathbf{z}\right)|^{2}+\sigma_{1}^{2}}\geq2^{\delta_{1}}.
&
\end{align}
Problem (10) is a nonconvex quadratically constrained quadratic program (QCQP). Therefore, we can relax it into a convex problem by using semidefinite relaxation (SDR). First, we define
\begin{align}
&
\mathbf{W}_{1}=\mathbf{w}_{1}\mathbf{w}_{1}^{H}, \nonumber
 \\
&
\mathbf{W}_{2}=\mathbf{w}_{2}\mathbf{w}_{2}^{H}, \nonumber
 \\
&
\mathbf{H}_{k}=\mathbf{G}^{H}\boldsymbol{\Phi}^{H}\mathbf{h}_{k}\mathbf{h}_{k}^{H}\boldsymbol{\Phi}\mathbf{G}, \nonumber
 \\
&
\alpha=\frac{|\mathbf{h}_{1}^{H}\boldsymbol{\Phi}\mathbf{G}\left(\mathbf{w}_{2}+\mathbf{z}\right)|^{2}+\sigma_{1}^{2}}{|\mathbf{h}_{e}^{H}\boldsymbol{\Phi}\mathbf{G}\left(\mathbf{w}_{2}+\mathbf{z}\right)|^{2}+\sigma_{e}^{2}}.
\end{align}
Based on the variable substitution, constraint C1a can be transformed into:
\begin{align}
&	\frac{\mathbf{\mathrm{Tr}}\left(\mathbf{H}_{1}\mathbf{W}_{1}\right)+\mathbf{\mathrm{Tr}}\left(\mathbf{H}_{1}\mathbf{W}_{2}\right)+\mathbf{\mathrm{Tr}}\left(\mathbf{H}_{1}\mathbf{Z}\right)+\sigma_{1}^{2}}{\mathbf{\mathrm{Tr}}\left(\mathbf{H}_{e}\mathbf{W}_{1}\right)+\mathbf{\mathrm{Tr}}\left(\mathbf{H}_{e}\mathbf{W}_{2}\right)+\mathbf{\mathrm{Tr}}\left(\mathbf{H}_{e}\mathbf{Z}\right)+\sigma_{e}^{2}}\geq2^{\delta_{1}}\alpha\nonumber
 \\
 &
\Leftrightarrow	\mathbf{\mathrm{Tr}}\left(\mathbf{H}_{1}\mathbf{W}_{1}\right)-2^{\delta_{1}}\alpha\mathbf{\mathrm{Tr}}\left(\mathbf{H}_{e}\mathbf{W}_{1}\right)+\beta\geq0,
&
\end{align}
where
\begin{align}
&
\beta=\mathrm{Tr}\left[\mathbf{H}_{1}\left(\mathbf{W}_{2}+\mathbf{Z}\right)\right]+\sigma_{1}^{2}\nonumber
 \\
 &
-2^{\delta_{1}}\alpha\left(\mathrm{Tr}\left[\mathbf{H}_{e}\left(\mathbf{W}_{2}+\mathbf{Z}\right)\right]+\sigma_{e}^{2}\right).
\end{align}
Constraint C1a becomes a convex semidefinite program (SDP) with respect to $\mathbf{W}_{1}$, which is more tractable.

Note that C2 is a nonconvex inequality constraint. Thus, it is likely that by
linearizing the nonconvex inequality with respect to $\mathbf{w}_{1}$,
C2 can be recast to
\begin{align}
\mathrm{C2a}:R_{g}= & \log_{2}\left(1+\frac{|\mathbf{h}_{2}^{H}\mathbf{\boldsymbol{\Phi}}\mathbf{G}\mathbf{w}_{2}|^{2}}{|\mathbf{h}_{2}^{H}\mathbf{\boldsymbol{\Phi}}\mathbf{G}\left(\mathbf{w}_{1}+\mathbf{z}\right)|^{2}+\sigma_{2}^{2}}\right)\geq\delta_{2}\nonumber \\
\Leftrightarrow & \frac{\mathbf{\mathrm{Tr}}\left(\mathbf{H}_{2}\mathbf{W}_{2}\right)}{\mathbf{\mathrm{Tr}}\left(\mathbf{H}_{2}\mathbf{W}_{1}\right)+\mathbf{\mathrm{Tr}}\left(\mathbf{H}_{2}\mathbf{Z}\right)+\sigma_{2}^{2}}\geq2^{\delta_{2}}-1\\
\Leftrightarrow & \mathbf{\mathrm{Tr}}\left(\mathbf{H}_{2}\mathbf{W}_{2}\right)-\left(2^{\delta_{2}}-1\right)\nonumber \\
 & \left(\mathbf{\mathrm{Tr}}\left(\mathbf{H}_{2}\mathbf{W}_{1}\right)+\mathbf{\mathrm{Tr}}\left(\mathbf{H}_{2}\mathbf{Z}\right)+\sigma_{2}^{2}\right)\geq0.\nonumber
\end{align}
Therefore, the new constraint C2a presents a convex linear matrix inequality (LMI) representation.

With the new constraints C1a and C2a, optimizing $\mathbf{w}_{1}$ in
$\mathbf{P1}$ can be reduced to
\begin{align}
\mathbf{P2:}\quad\underset{\mathbf{W}_{1}}{\min}\; & \mathbf{\mathrm{Tr}}\left(\mathbf{W}_{1}\right)\nonumber \\
st.\; & \mathrm{C1a}:\mathbf{\mathrm{Tr}}\left(\mathbf{H}_{1}\mathbf{W}_{1}\right)-2^{\delta_{1}}\alpha\mathbf{\mathrm{Tr}}\left(\mathbf{H}_{e}\mathbf{W}_{1}\right)+\beta\geq0,\nonumber \\
 & \mathrm{C2a}:\mathbf{\mathrm{Tr}}\left(\mathbf{H}_{2}\mathbf{W}_{2}\right)-\left(2^{\delta_{2}}-1\right)\\
 & \left(\mathbf{\mathrm{Tr}}\left(\mathbf{H}_{2}\mathbf{W}_{1}\right)+\mathbf{\mathrm{Tr}}\left(\mathbf{H}_{2}\mathbf{Z}\right)+\sigma_{2}^{2}\right)\geq0,\nonumber \\
 & \mathrm{C5}:\mathbf{W}_{1}\succeq0,\nonumber
\end{align}
and constraint C5 is imposed to guarantee $\mathbf{W}_{1}=\mathbf{w}_{1}\mathbf{w}_{1}^{H}$. Now, problem $\mathbf{P2}$ is convex in auxiliary matrix $\mathbf{W}_{1}$ and can be solved by CVX. If the solution $\mathbf{W}_{1}$ satisfies the rank-one constraint, the optimal beamformer $\mathbf{w}_{1}$ can be recovered from $\mathbf{W}_{1}$ by performing the classic Cholesky decomposition \cite{nocedal2006numerical}. However, in general, the solution obtained by SDR cannot exhibit the rank-one property. We can find the suboptimal $\mathbf{w}_{1}$ by applying the rank-one approximations of $\mathbf{W}_{1}$, such as eigenvalue decomposition (EVD) with Gaussian randomization. To be specific, we decompose $\mathbf{W}_{1}$ as $\ensuremath{\mathbf{W}_{1}}=\mathbf{U}_{v}\boldsymbol{\Sigma}_{v}\mathbf{U}_{v}^{H}$, where $\mathbf{U}_{v}=\left[u_{v,1},...u_{v,M}\right]$ and $\boldsymbol{\Sigma}_{v}=\mathrm{diag}\left(\lambda_{v,1},...,\lambda_{v,M}\right)$ are a unitary matrix and a diagonal matrix, respectively. We can generate a vector as $\mathbf{\overline{w}}_{1}=\mathbf{U}_{v}\boldsymbol{\Sigma}_{v}^{1/2}\boldsymbol{r}_{l}$ where $\boldsymbol{r}_{l}\sim\mathrm{\mathcal{CN}}\left(0,\boldsymbol{I}_{M}\right)$ and $\boldsymbol{r}_{l}$ is the randomly generated vector satisfying Gaussian distribution with zero means. Then, we can obtain a suboptimal solution from multiple approximate solutions. It should be noted that the SDR is an efficient and widely used convex approximation approach, which can achieve satisfactory performance and approach the optimal performance \cite{9133130,9310230}.

\subsection{Optimization of the Beamformer $\mathbf{w}_{2}$ and the AN $\mathbf{z}$}

Due to the fact that the algebraic expressions involved in the optimization
problem are in the symmetric functions $\mathbf{w}_{2}$ and $\mathbf{z}$,
we can jointly optimize the general beamformer $\mathbf{w}_{2}$
and the AN $\mathbf{z}$. Given the confidential beamformer $\mathbf{w}_{1}$
and the reflection phase-shifts $\boldsymbol{\Phi}$ of the IRS, optimizing
$\mathbf{w}_{2}$ and $\mathbf{z}$ involve only constraints
C1, C2 and C3; therefore, C4 can be omitted.

Now, for a given $\mathbf{w}_{1}$ and $\boldsymbol{\Phi}$, we can recast
C1 as follows:

\begin{align}
\mathrm{C1b}:&R_{sec}=\log_{2}\left(1+\frac{|\mathbf{h}_{1}^{H}\boldsymbol{\Phi}\mathbf{G}\mathbf{w}_{1}|^{2}}{|\mathbf{h}_{1}^{H}\boldsymbol{\Phi}\mathbf{G}\left(\mathbf{w}_{2}+\mathbf{z}\right)|^{2}+\sigma_{1}^{2}}\right)\nonumber
\\
&
-\log_{2}\left(1+\frac{|\mathbf{h}_{e}^{H}\boldsymbol{\Phi}\mathbf{G}\mathbf{w}_{1}|^{2}}{|\mathbf{h}_{e}^{H}\boldsymbol{\Phi}\mathbf{G}\left(\mathbf{w}_{2}+\mathbf{z}\right)|^{2}+\sigma_{e}^{2}}\right)\geq\delta_{1}\\
  &=-N_{1}+N_{2}+N_{3}-N_{4}\geq\delta_{1},\nonumber
\end{align}
where

\begin{align}
N_{1} =&-\log_{2}\left(\mathrm{Tr}\left(\mathbf{H}_{1}\mathbf{W}_{2}\right)+\mathrm{Tr}\left(\mathbf{H}_{1}\mathbf{Z}\right)\right.\nonumber
\\
&\left.+\mathrm{Tr}\left(\mathbf{H}_{1}\mathbf{W}_{1}\right)+\sigma_{1}^{2}\right),
\\
N_{2}  = &-\log_{2}\left(\mathrm{Tr}\left(\mathbf{H}_{1}\mathbf{W}_{2}\right)+\mathrm{Tr}\left(\mathbf{H}_{1}\mathbf{Z}\right)+\sigma_{1}^{2}\right),
\\
N_{3} =&-\log_{2}\left(\mathrm{Tr}\left(\mathbf{H}_{e}\mathbf{W}_{2}\right)+\mathrm{Tr}\left(\mathbf{H}_{e}\mathbf{Z}\right)\right.\nonumber
\\
&\left.+\mathrm{Tr}\left(\mathbf{H}_{e}\mathbf{W}_{1}\right)+\sigma_{e}^{2}\right),\\
N_{4} =&-\log_{2}\left(\mathrm{Tr}\left(\mathbf{H}_{e}\mathbf{W}_{2}\right)+\mathrm{Tr}\left(\mathbf{H}_{e}\mathbf{Z}\right)+\sigma_{e}^{2}\right).
\end{align}
Note that $N_{2}$ and $N_{3}$ are in convex form with
respect to $\mathbf{w}_{2}$ and $\mathbf{z}$ in (16), but
the terms $-N_{1}$ and $-N_{4}$ are in concave form with respect
to $\mathbf{w}_{2}$ and $\mathbf{z}$. Hence, (16) is nonconvex due
to the difference-of-convex form \cite{9133130}. To address the nonconvexity,
we resort to successive convex approximation (SCA) to construct a
global underestimator for the function $N_{1}$ and $N_{4}$. Their first-order
Taylor approximation expressions, which can be written as discussed in (\ref{eq:the expand of N1})
and (\ref{eq:the expand of N4}), are shown at the top of next page.
\begin{figure*}
\begin{align}
N_{1}\left(\mathbf{W}_{2},\mathbf{Z}\right) & \geq N_{1}\left(\mathbf{W}_{2}^{\left(t\right)},\mathbf{Z}^{\left(t\right)}\right)+\mathrm{Tr}\left(\nabla_{\mathbf{W}_{2}}N_{1}\left(\mathbf{W}_{2}^{\left(t\right)},\mathbf{Z}^{\left(t\right)}\right)\left(\mathbf{W}_{2}-\mathbf{W}_{2}^{\left(t\right)}\right)+\mathrm{Tr}\left(\nabla_{\mathbf{Z}}N_{1}\left(\mathbf{W}_{2}^{\left(t\right)},\mathbf{Z}^{\left(t\right)}\right)\left(\mathbf{Z}-\mathbf{\mathbf{Z}}^{\left(t\right)}\right)\right)\right),\label{eq:the expand of N1}\\
N_{4}\left(\mathbf{W}_{2},\mathbf{Z}\right) & \geq N_{4}\left(\mathbf{W}_{2}^{\left(t\right)},\mathbf{Z}^{\left(t\right)}\right)+\mathrm{Tr}\left(\nabla_{\mathbf{W}_{2}}N_{4}\left(\mathbf{W}_{2}^{\left(t\right)},\mathbf{Z}^{\left(t\right)}\right)\left(\mathbf{W}_{2}-\mathbf{W}_{2}^{\left(t\right)}\right)+\mathrm{Tr}\left(\nabla_{\mathbf{Z}}N_{4}\left(\mathbf{W}_{2}^{\left(t\right)},\mathbf{Z}^{\left(t\right)}\right)\left(\mathbf{Z}-\mathbf{\mathbf{Z}}^{\left(t\right)}\right)\right)\right).\label{eq:the expand of N4}
\end{align}

\hrulefill
\end{figure*}
In (\ref{eq:the expand of N1}) and (\ref{eq:the expand of N4}),
\begin{align}
 & \nabla_{\mathbf{W}_{2}}N_{1}\left(\mathbf{W}_{2},\mathbf{Z}\right)=\nabla_{\mathbf{Z}}N_{1}\left(\mathbf{W}_{2},\mathbf{Z}\right) \\
 & =-\frac{1}{\ln2}\frac{\mathbf{H}_{1}}{\mathrm{Tr}\left(\mathbf{H}_{1}\mathbf{W}_{2}\right)+\mathrm{Tr}\left(\mathbf{H}_{1}\mathbf{Z}\right)+\mathrm{Tr}\left(\mathbf{H}_{1}\mathbf{W}_{1}\right)+\sigma_{1}^{2}},\nonumber\\
 & \nabla_{\mathbf{W}_{2}}N_{4}\left(\mathbf{W}_{2},\mathbf{Z}\right)=\nabla_{\mathbf{Z}}N_{4}\left(\mathbf{W}_{2},\mathbf{Z}\right)\nonumber \\
 & =-\frac{1}{\ln2}\frac{\mathbf{H}_{e}}{\mathrm{Tr}\left(\mathbf{H}_{e}\mathbf{W}_{2}\right)+\mathrm{Tr}\left(\mathbf{H}_{e}\mathbf{Z}\right)+\sigma_{e}^{2}}.
\end{align}
Then, we can replace the nonconvex constraint C2 with the tractable expression
C2a in (14).

Based on the above the transformations, optimizing $\mathbf{w}_{2}$
and $\mathbf{z}$ in $\mathbf{P1}$ is reduced to

\begin{align}
\mathbf{P3:}\qquad & \underset{\mathbf{W}_{2},\mathbf{Z}}{\min}\;\;\mathbf{\mathrm{Tr}}\left(\mathbf{W}_{2}\right)+\mathbf{\mathrm{Tr}}\left(\mathbf{Z}\right)\nonumber \\
s.t.\qquad & \mathrm{C1b}:N_{2}+N_{3}\nonumber \\
 & -\mathrm{Tr}\left(\nabla_{\mathbf{W}_{2}}N_{1}\left(\mathbf{W}_{2}^{\left(t\right)},\mathbf{Z}^{\left(t\right)}\right)\mathbf{W}_{2}\right)\nonumber \\
 & -\mathrm{Tr}\left(\nabla_{\mathbf{Z}}N_{1}\left(\mathbf{W}_{2}^{\left(t\right)},\mathbf{Z}^{\left(t\right)}\right)\mathbf{Z}\right)\nonumber \\
 & -\mathrm{Tr}\left(\nabla_{\mathbf{W}_{2}}N_{4}\left(\mathbf{W}_{2}^{\left(t\right)},\mathbf{Z}^{\left(t\right)}\right)\mathbf{W}_{2}\right)\nonumber \\
 & -\mathrm{Tr}\left(\nabla_{\mathbf{Z}}N_{4}\left(\mathbf{W}_{2}^{\left(t\right)},\mathbf{Z}^{\left(t\right)}\right)\mathbf{Z}\right)\geq\delta_{1},\nonumber \\
 & \mathrm{C2a}:\mathbf{\mathrm{Tr}}\left(\mathbf{W}_{2}\mathbf{H}_{2}\right)-\left(2^{\delta_{2}}-1\right)\\
 & \left(\mathbf{\mathrm{Tr}}\left(\mathbf{W}_{1}\mathbf{H}_{2}\right)+\mathbf{\mathrm{Tr}}\left(\mathbf{Z}\mathbf{H}_{2}\right)+\sigma_{2}^{2}\right)\geq0,\nonumber \\
 & \mathrm{C3}:\mathbf{Z}\succeq0,\nonumber \\
 & \mathrm{C6}:\mathbf{W}_{2}\succeq0,\nonumber
\end{align}
where C6 is introduced to interpret the relaxation of the constraint
$\mathbf{W}_{2}=\mathbf{w}_{2}\mathbf{w}_{2}^{H}$
during the application of the SDR method. For the convex problem $\mathbf{P3}$, the EVD with Gaussian randomization can be applied to recover the optimal $\mathbf{w}_{2}$ and $\mathbf{z}$ after the solutions $\mathbf{W}_{2}$ and $\mathbf{Z}$ are obtained similar to restoring $\mathbf{w}_{1}$.

\subsection{Optimization of the IRS Phase Shifts $\boldsymbol{\boldsymbol{\Phi}}$}

In this system, by reconfiguring the wireless environments, IRS is
introduced to bring additional beamforming gains in the expected
directions while suppressing undesirable interference. Therefore,
the passive beamforming gain can theoretically help save active transmit power
at the BS. Given the beamformers $\mathbf{w}_{1},\mathbf{w}_{2}$
and the AN vector $\mathbf{z}$, only constraints C1, C2 and C4 are
involved in the optimization of $\mathbf{P1}$. The main difficulty
in solving the passive beamforming subproblem lies in the highly intractable
form of the nonconvex unit modulus constraint C4.

By changing the variables $\mathbf{h}_{k}^{H}\boldsymbol{\Phi}\mathbf{G}=\boldsymbol{v}^{H}\boldsymbol{\varTheta}_{k}$
to $\boldsymbol{\varTheta}_{k}=\mathrm{diag}\left(\mathbf{h}_{k}^{H}\right)\mathbf{G}$,
we can write C1 as follows:

\begin{align}
\mathrm{C1c}:R_{sec} & =\log_{2}\left(1+\frac{|\boldsymbol{v}^{H}\boldsymbol{\varTheta}_{1}\mathbf{w}_{1}|^{2}}{|\boldsymbol{v}^{H}\boldsymbol{\varTheta}_{1}\left(\mathbf{w}_{2}+\mathbf{z}\right)|^{2}+\sigma_{1}^{2}}\right) \\
 & -\log_{2}\left(1+\frac{|\boldsymbol{v}^{H}\boldsymbol{\varTheta}_{e}\mathbf{w}_{1}|^{2}}{|\boldsymbol{v}^{H}\boldsymbol{\varTheta}_{e}\left(\mathbf{w}_{2}+\mathbf{z}\right)|^{2}+\sigma_{e}^{2}}\right)\geq\delta_{1}.\nonumber
\end{align}
However, analyzing the quadratic terms in (26) is still challenging. To facilitate the SDR form, we define $\mathbf{M}_{k,i}=\mathbf{w}_{i}^{H}\boldsymbol{\varTheta}_{k}^{H}\boldsymbol{\varTheta}_{k}\mathbf{w}_{i}$,
$\mathbf{M}_{k,z}=\mathbf{z}^{H}\boldsymbol{\varTheta}_{k}^{H}\boldsymbol{\varTheta}_{k}\mathbf{z}$,
$\boldsymbol{O}_{k}=\boldsymbol{\varTheta}_{k}^{H}\boldsymbol{\varTheta}_{k}$,
and $\boldsymbol{V}=\boldsymbol{v}\boldsymbol{v}^{H}$. Then, we can
rewrite $R_{sec}$ in (26) as

\begin{align}
R_{sec} & =\log_{2}\left(P_{t}\mathrm{Tr}\left(\boldsymbol{V}\boldsymbol{O}_{1}\right)+\sigma_{1}^{2}\right)\nonumber \\
 & -\log_{2}\left(\mathrm{Tr}\left(\boldsymbol{V}\mathbf{M}_{1,2}\right)+\mathrm{Tr}\left(\boldsymbol{V}\mathbf{M}_{1,z}\right)+\sigma_{1}^{2}\right)\nonumber \\
 & -\log_{2}\left(P_{t}\mathrm{Tr}\left(\boldsymbol{V}\boldsymbol{O}_{e}\right)+\sigma_{e}^{2}\right)\nonumber \\
 & +\log_{2}\left(\mathrm{Tr}\left(\boldsymbol{V}\mathbf{M}_{e,2}\right)+\mathrm{Tr}\left(\boldsymbol{V}\mathbf{M}_{e,z}\right)+\sigma_{e}^{2}\right)\geq\delta_{1}\nonumber \\
 & =-T_{1}+T_{2}+T_{3}-T_{4}\geq\delta_{1},
\end{align}
where
\begin{align}
T_{1} & =-\log_{2}\left(P_{t}\mathrm{Tr}\left(\boldsymbol{V}\boldsymbol{O}_{1}\right)+\sigma_{1}^{2}\right),\\
T_{2} & =-\log_{2}\left(\mathrm{Tr}\left(\boldsymbol{V}\mathbf{M}_{1,2}\right)+\mathrm{Tr}\left(\boldsymbol{V}\mathbf{M}_{1,z}\right)+\sigma_{1}^{2}\right),\\
T_{3} & =-\log_{2}\left(P_{t}\mathrm{Tr}\left(\boldsymbol{V}\boldsymbol{O}_{e}\right)+\sigma_{e}^{2}\right),
\end{align}
and
\begin{equation}
T_{4}  =-\log_{2}\left(\mathrm{Tr}\left(\boldsymbol{V}\mathbf{M}_{e,2}\right)+\mathrm{Tr}\left(\boldsymbol{V}\mathbf{M}_{e,z}\right)+\sigma_{e}^{2}\right).
\end{equation}

Because $T_{2}$ and $T_{3}$ are in convex form with $\boldsymbol{V}$
while the terms $T_{1}$ and $T_{4}$ are in concave form with respect
to $\boldsymbol{V}$, $R_{sec}$ in (27) is a typical difference-of-convex
form. To approximate this nonconvex form, the first-order
Taylor approximation in (28) and (31) is applied, which can be written as

\begin{align}
T_{1}\left(\boldsymbol{V}\right) & \geq T_{1}\left(\boldsymbol{V}^{\left(t\right)}\right)+\mathrm{Tr}\left(\nabla_{\boldsymbol{V}}T_{1}\left(\boldsymbol{V}^{\left(t\right)}\right)\left(\boldsymbol{V}-\boldsymbol{V}^{\left(t\right)}\right)\right),\\
T_{4}\left(\boldsymbol{V}\right) & \geq T_{4}\left(\boldsymbol{V}^{\left(t\right)}\right)+\mathrm{Tr}\left(\nabla_{\boldsymbol{V}}T_{4}\left(\boldsymbol{V}^{\left(t\right)}\right)\left(\boldsymbol{V}-\boldsymbol{V}^{\left(t\right)}\right)\right),
\end{align}
where
\begin{align}
\nabla_{\boldsymbol{V}}T_{1}\left(\boldsymbol{V}\right) & =-\frac{1}{\ln2}\frac{\boldsymbol{O}_{1}}{P\mathrm{Tr}\left(\boldsymbol{V}\boldsymbol{O}_{1}\right)+\sigma_{1}^{2}},\\
\nabla_{\boldsymbol{V}}T_{4}\left(\boldsymbol{V}\right) & =-\frac{1}{\ln2}\frac{\mathbf{M}_{e,2}+\mathbf{M}_{e,z}}{\mathrm{Tr}\left(\boldsymbol{V}\mathbf{M}_{e,2}\right)+\mathrm{Tr}\left(\boldsymbol{V}\mathbf{M}_{e,z}\right)+\sigma_{e}^{2}}.
\end{align}
Accordingly, the secure rate $R_{sec}$ can be updated as follows:
\begin{align}
R_{sec} & =-T_{1}+T_{2}+T_{3}-T_{4}\nonumber \\
 & =T_{2}+T_{3}-\mathrm{Tr}\left(\nabla_{\boldsymbol{V}}T_{1}\left(\boldsymbol{V}^{\left(t\right)}\right)\boldsymbol{V}\right)\\
 & -\mathrm{Tr}\left(\nabla_{\boldsymbol{V}}T_{4}\left(\boldsymbol{V}^{\left(t\right)}\right)\boldsymbol{V}\right)\geq\delta_{1}.\nonumber
\end{align}

In a similar manner as was done in (26), C2 can be recast to
\begin{align}
\mathrm{C2b}:R_{e} & =\log_{2}(1+\frac{|\boldsymbol{v}^{H}\boldsymbol{\varTheta}_{2}\mathbf{w}_{2}|^{2}}{|\boldsymbol{v}^{H}\boldsymbol{\varTheta}_{2}\left(\mathbf{w}_{1}+\mathbf{z}\right)|^{2}+\sigma_{2}^{2}})\geq\delta_{2}\nonumber \\
 & \Leftrightarrow\mathrm{Tr}\left(\boldsymbol{V}\mathbf{M}_{2,2}\right)-\left(2^{\delta_{2}}-1\right)\\
 & \left(\mathrm{Tr}\left[\boldsymbol{V}\left(\mathbf{M}_{2,1}+\mathbf{M}_{2,z}\right)\right]\right)\geq0.\nonumber
\end{align}

After the above transformations, optimizing the reflection coefficients
in problem $\mathbf{P1}$ can be reduced to

\begin{align}
\mathbf{P4:}\qquad & \underset{\boldsymbol{V}}{\mathrm{Find}}\;\;\boldsymbol{V}\nonumber \\
s.t.\qquad & \mathrm{C1c}:T_{2}+T_{3}-\mathrm{Tr}\left(\nabla_{\boldsymbol{V}}T_{1}\left(\boldsymbol{V}^{\left(t\right)}\right)\boldsymbol{V}\right)\nonumber \\
 & -\mathrm{Tr}\left(\nabla_{\boldsymbol{V}}T_{4}\left(\boldsymbol{V}^{\left(t\right)}\right)\boldsymbol{V}\right)\geq\delta_{1},\nonumber \\
 & \mathrm{C2b}:\mathrm{Tr}\left(\boldsymbol{V}\mathbf{M}_{2,2}\right)-\left(2^{\delta_{2}}-1\right)\\
 & \left(\mathrm{Tr}\left(\boldsymbol{V}\mathbf{M}_{2,1}\right)+\mathrm{Tr}\left(\boldsymbol{V}\mathbf{M}_{2,z}\right)\right)\geq0,\nonumber \\
 & \mathrm{C7}:\boldsymbol{V}\succeq0,\nonumber \\
 & \mathrm{C8}:\mathrm{Diag}\left(\boldsymbol{V}\right)=1_{N},\nonumber
\end{align}
where C7 is newly added to relax the equation constraint $\boldsymbol{V}=\boldsymbol{v}\boldsymbol{v}^{H}$
during the SDR procedures, and C8 is introduced to guarantee the unit
modulus constraint.

Problem $\mathbf{P4}$ is now in convex form and hence is solvable. Given the variables $\boldsymbol{V}^{*}$ obtained by the CVX tools, we then employ the EVD with Gaussian randomization so the reflection phase shift matrix $\boldsymbol{\Phi}$ can be restored similar to restoring $\mathbf{w}_{1}$.

To simplify the understanding of the proposed alternating optimization framework,
the overall optimization procedures of $\mathbf{w}_{1},\mathbf{w}_{2},\mathbf{z}$
and $\boldsymbol{\Phi}$ are summarized in Algorithm 1. To illustrate
the convergence of the proposed approach, we provide the corresponding
proof in Proposition 1.

\emph{Proposition 1:} Algorithm 1 is guaranteed to converge when the
proposed alternating iterative framework is used.
\begin{IEEEproof}
See Appendix A.
\end{IEEEproof}
Moreover, we provide a computational complexity analysis of the proposed algorithm.  As described in Algorithm 1, the complexity is dominated by steps (a), (b) and (c). To be specific, running the SDP in (a) and (b) results in the complexity of $\mathcal{O}\left(N_{A}^{4.5}\right)$ and $\mathcal{O}\left(N_{B}^{4.5}\right)$ \cite{Polik2010interior}, respectively. Solving (c) for the reflection coefficient requires the complexity of $\mathcal{O}\left(\left(N+1\right)^{4.5}\right)$ \cite{8723525}. The overall complexity of Algorithm 1 is $\mathcal{O}\left(L\left(N_{A}^{4.5}+N_{B}^{4.5}+\left(N+1\right)^{4.5}\right)\right)$, where $L$ denotes the iteration number of the algorithm.

\begin{algorithm}
\begin{enumerate}
\item Initialization: Set the iteration index to $t=1$, construct the general
beamformer $\mathbf{w}_{2}^{*}=\sqrt{P_{t}^{*}}\frac{\mathbf{\left(\mathbf{h}_{2}^{\mathit{H}}\mathbf{\boldsymbol{\Phi}}\mathbf{G}\right)}^{H}}{\Vert\mathbf{h}_{2}^{H}\mathbf{\boldsymbol{\Phi}}\mathbf{G}\Vert}$,
and generate the initial AN vector $\mathbf{z}$ and initial IRS coefficient
matrix $\boldsymbol{\Phi}$.
\item Repeat
\begin{enumerate}
\item Obtain the confidential beamformer $\mathbf{w}_{1}^{\left(t\right)}$
by solving $\mathbf{P2}$ with the given variables $\mathbf{w}_{2}^{*},\mathbf{z}$
and $\boldsymbol{\Phi}$.
\item Obtain the general beamformer $\mathbf{w}_{2}^{\left(t\right)}$ and
the AN $\mathbf{z}^{\left(t\right)}$ by solving $\mathbf{P3}$ with
the fixed $\boldsymbol{\Phi}$ and $\mathbf{w}_{1}^{\left(t\right)}$.
\item Find the reflection coefficient matrix $\boldsymbol{\Phi}^{\left(t\right)}$
by solving $\mathbf{P4}$ with the fixed $\mathbf{w}_{1}^{\left(t\right)},\mathbf{w}_{2}^{\left(t\right)}$
and $\mathbf{z}^{\left(t\right)}$.
\item Substitute the updated variables $\mathbf{w}_{1}^{\left(t\right)},\mathbf{w}_{2}^{\left(t\right)}$
and $\mathbf{z}^{\left(t\right)}$ into the original problem to obtain
the power $P_{t}^{\left(t\right)}$.
\item Update $t=t+1.$
\end{enumerate}
\item Until $|\frac{P_{t}^{(t)}-P_{t}^{(t+1)}}{P_{t}^{(t+1)}}|<\delta$
or the maximum number of iterations is reached. Then, terminate the algorithm.
\end{enumerate}
\caption{Alternating Algorithm-based Transmit Power Minimization.}
\end{algorithm}

\subsection{The execution entity of the proposed scheme}
Initially, the BS performs channel estimation through the pilots sent by users and collects the channel state information of all involved users. Next, the users write the required different security requests into the data packet and transmit it to the BS. Then, the BS receives the service request of the users and divides it into confidential subscribers or general communication users according to the WAP. Finally, by running the proposed algorithm at the BS, we can calculate the transmit beamforming vectors, the injected AN vector and the reflection phase shifts. The beamforming design and artificial noise superposition design are carried out for the transmitted signal at the BS. The IRS controller adjusts the reflection amplitude and angle through the instructions sent by the BS. Through the cooperation of the BS, the IRS and UEs, the downlink secure transmission scheme is ensured.

\section{Simulation Results}

In this section, we evaluate the performance of the proposed algorithm
in an IRS-aided security classification wireless communication system.

\subsection{Simulation Settings}

\begin{figure}
\centering{}\includegraphics[scale=0.5]{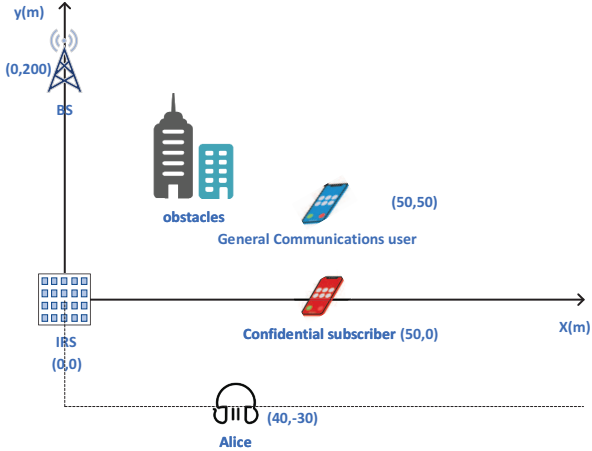}

\caption{The simulated IRS-aided security-classification wireless communication
scenario.}
\end{figure}

The downlink transmission simulation setup is depicted in Fig. 2,
where the radius is three hundred meters. The coordinate of the IRS
and the BS are denoted by $\left(0,0\right)$ and $\left(0,200\right)$,
respectively. The confidential subscriber, the general communication
user and Alice are located at $\left(50,0\right),\left(50,50\right)$
and $\left(40,-30\right)$, respectively. We assume that
obstacles exist between the BS and the users; therefore, the signals can be
transmitted to the users only through the BS--IRS--User link. The channel matrix
$\mathbf{G}$ between the BS and the IRS follows a Rician distribution
because it is often deployed in open areas with good channel quality.
The channel matrix $\mathbf{G}$ is structured as follows:

\begin{equation}
\mathbf{G}=\sqrt{\beta}\left(\sqrt{\frac{\delta}{\delta+1}}\bar{\mathbf{G}}+\sqrt{\frac{\delta}{\delta+1}}\tilde{\mathbf{G}}\right),
\end{equation}
where $\beta$ is the large-scale fading coefficient, $\delta$ is the
Rician factor, and $\tilde{\mathbf{G}}$ is the non-line-of-sight (NLoS)
channel component whose elements are independent and identically distributed
with $\mathrm{\mathcal{CN}}\left(0,1\right)$ \cite{9279316}.

According to the uniform square planar array (USPA) model, $\bar{\mathbf{G}}$
represents the line-of-sight (LoS) channel component, as given by
\begin{equation}
\bar{\mathbf{G}}=e^{-j2\pi\frac{d}{\lambda}}\mathbf{a}_{I}\left(\psi_{a},\psi_{e}\right)\mathbf{a}_{B}^{H}\left(\psi_{b}\right),
\end{equation}
where $d$ is the distance between the BS and the IRS. $\lambda$
is the wavelength. $\psi_{a}$ and $\psi_{e}$ are the azimuth and elevation
angles of arrival at the IRS, respectively. $\psi_{b}$ is the angle
of departure from the BS to the IRS. $\mathbf{a}_{I}$ and $\mathbf{a}_{B}$ are the steering vectors corresponding to the IRS and BS, respectively.

The channel matrix $\mathbf{h}_{k}$ between the IRS and the $k$-th
user follows the the Rayleigh distribution, and $\alpha$ is the path
loss exponent. $t$ is the maximum number of iterations, and $\epsilon$
is the error accuracy. The main parameters used in our simulations are
provided in Table II.

\begin{table}[b]
\caption{SIMULATION PARAMETERS }

\centering{}%
\begin{tabular}{|c|c|}
\hline
Carrier center frequency & 2.4 GHz\tabularnewline
\hline
Distance between BS and IRS $d$ & 200 m\tabularnewline
\hline
Path loss exponent $\alpha$ & 4\tabularnewline
\hline
Path loss $\beta$ & $10^{-9}$\tabularnewline
\hline
Rician factor $\delta$ & 10\tabularnewline
\hline
Cell radius $R_{c}$ & 300 m\tabularnewline
\hline
Noise power & -90 dBm\tabularnewline
\hline
Maximum number of iterations $t$ & 1000\tabularnewline
\hline
Error accuracy $\epsilon$ & $10^{-4}$\tabularnewline
\hline
\end{tabular}
\end{table}

\subsection{Performance Evaluation}

To validate the performance of the proposed scheme, we selected four baseline
schemes for performance comparisons. The considered baseline
designs are as follows:
\begin{itemize}
\item \textbf{Baseline 1:} We derive the confidential beamformer and the
general beamformer according to the maximum ratio transmission (MRT)
based scheme, i.e., $\mathbf{w}_{1}^{*}=\sqrt{P_{t}^{*}}\frac{\mathbf{\left(\mathbf{h}_{1}^{\mathit{H}}\mathbf{\boldsymbol{\Phi}}\mathbf{G}\right)}^{H}}{\Vert\mathbf{h}_{1}^{H}\mathbf{\boldsymbol{\Phi}}\mathbf{G}\Vert}$
and $\mathbf{w}_{2}^{*}=\sqrt{P_{t}^{*}}\frac{\mathbf{\left(\mathbf{h}_{2}^{\mathit{H}}\mathbf{\boldsymbol{\Phi}}\mathbf{G}\right)}^{H}}{\Vert\mathbf{h}_{2}^{H}\mathbf{\boldsymbol{\Phi}}\mathbf{G}\Vert}$.
Then, we apply Algorithm 1 without further updating $\mathbf{w}_{1}$ and
$\mathbf{w}_{2}$ to obtain the other two: $\mathbf{z}$ and
$\boldsymbol{\Phi}$. This baseline is termed the ``MRT-based scheme''.
\item \textbf{Baseline 2:} We randomly select the reflection phase shifts
$\boldsymbol{\Phi}$ at the IRS, and the other three variables $\mathbf{w}_{1},\mathbf{w}_{2}$.
Then, $\mathbf{z}$ is attained by applying Algorithm 1 without updating
the $\boldsymbol{\Phi}$. This baseline is termed the ``Rand phase-based
scheme''.
\item \textbf{Baseline 3:} We obtain the beamformers $\mathbf{w}_{1},\mathbf{w}_{2}$
and the reflection phase shifts $\boldsymbol{\Phi}$ using Algorithm
1 but remove the procedure to inject the AN at the BS. This baseline is termed the ``w/o AN-based scheme''.
\item \textbf{Baseline 4:} We adopt the traditional relay based scheme without
IRS, which forwards the confidential information via amplification.
This baseline is termed the ``Relay-based scheme''.
\end{itemize}

\begin{figure}
\centering{}\includegraphics[scale=0.6]{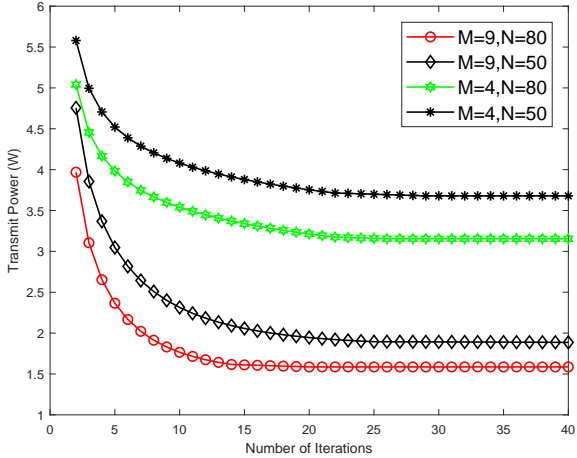}

\caption{Convergence behavior of the proposed algorithm.}
\end{figure}

As shown in Fig. 3, we investigated the convergence behavior of the proposed algorithm. Under the scenarios of different antenna numbers and IRS numbers, the proposed algorithm exhibits good convergence after a certain number of iterations. As the number of transmit antennas and the reflecting elements increase, the transmit power reduces accordingly. By maintaining the number of transmit antennas and reducing the number of IRS reflecting elements, the system transmit power decreases by only $10\%$. Conversely, when the number of the IRS reflecting elements is fixed, reducing the number of the transmit antennas can decrease the required transmit power by $30\%$. The result shows that the number of transmit antennas contributes more to saving active power consumption than does the IRS.

\begin{figure}[t]
\centering{}\subfigure[Alice is located at (40,-30), and the confidential subscriber is moving away from the IRS.]{\includegraphics[scale=0.6]{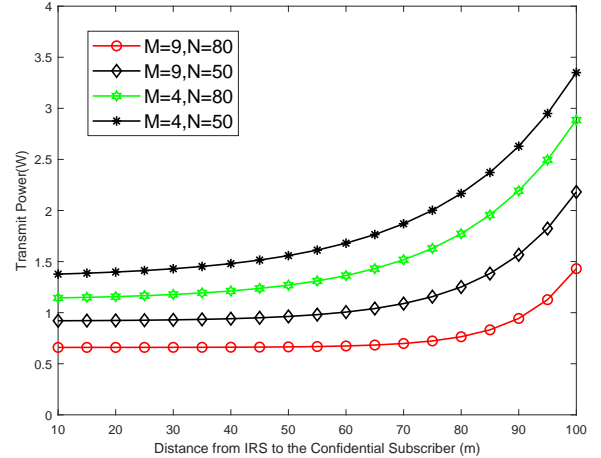}}
\subfigure[The confidential subscriber is located at (50,0), and Alice is moving away from the IRS.]{\includegraphics[scale=0.6]{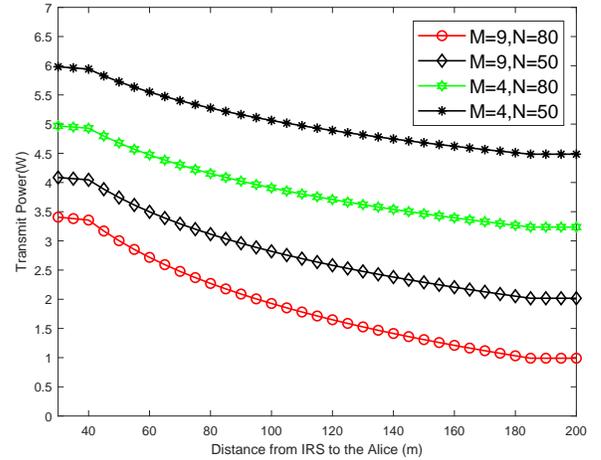}}

\caption{Transmit power versus communication users' distance (m).}
\end{figure}

In Fig. 4(a), we study the influence of the distance between the confidential
subscriber and the IRS on the system transmit power. In our setup,
the positions of the general communication user and Alice in the system
are fixed at $\left(50,50\right)$ and $\left(40,-30\right)$, respectively. The confidential subscriber gradually moves from $\left(10,0\right)$
to $\left(200,0\right)$ along the $x$-axis. It can be seen that as the
distance between the confidential subscriber and the IRS increases,
the BS transmit power must increase to ensure
the security rate requirements of the communication user. Meanwhile,
adding more transmit antennas or reflection units at the IRS can reduce the transmit power effectively.

Fig. 4(b) evaluates the impact of the distance between Alice
and the IRS on the system transmit power. The positions
of the general communication user and the confidential subscriber
are fixed at $\left(50,50\right)$ and $\left(50,0\right)$, respectively. Alice
gradually moves from $\left(0,-30\right)$ to $\left(200,-30\right)$ along the dotted
line shown in Fig. 4(b). As Alice started to stay away from the IRS, the required transmit power decreases slowly. In contrast, as Alice moves to $\left(40,-30\right)$ and then outward,
the required transmit power decreases faster. This is because Alice is farther away from the IRS than the confidential
subscriber. When moving to $\left(185,-30\right)$, Alice is sufficiently
far from the coverage of the BS. The required transmit
power remains almost unchanged as Alice proceeds to move
outward, and it can be considered that no
eavesdroppers are present in the system.

\begin{figure}
\centering{}\includegraphics[scale=0.6]{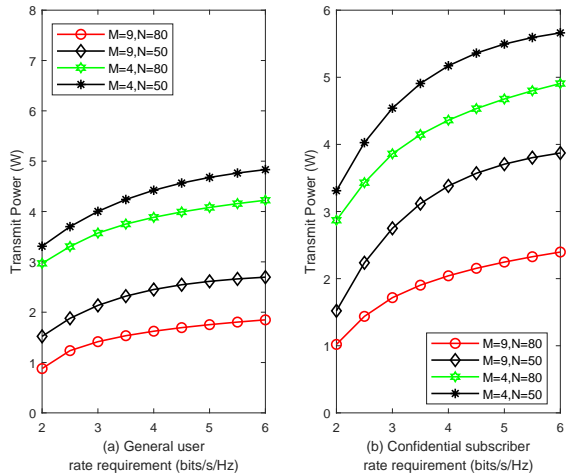}

\caption{System transmit power versus the rate requirement (bit/s/Hz).}
\end{figure}

Fig. 5 plots the system transmit power versus the communication
rate requirements. In Fig. 5(a), the minimum
communication rate of the confidential subscriber is set to 4
bits/s/Hz. As the minimum transmit rate of the general communication user increases, the transmit power increases slowly.
In Fig. 5(b), we set the minimum transmit rate of the general
communication user to 4 bits/s/Hz. As the minimum transmit
rate of the confidential subscriber gradually increases, the BS
transmit power increases rapidly. Under the same rate requirement, the confidential subscriber requires more power (which is more expensive) compared with the general communication user to transmit information.

\begin{figure}[t]
\centering{}\includegraphics[scale=0.45]{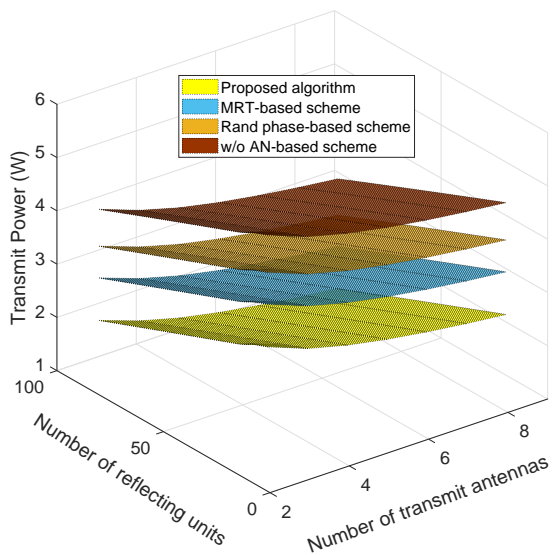}\caption{System transmit power versus the numbers of transmit antennas and reflecting units.}
\end{figure}

Fig. 6 compares the performances of the proposed algorithm and three
baseline algorithms under different numbers of transmit antennas and
reflecting units. The minimum communication rate requirements of the
confidential subscriber and the general communication user are fixed
to $4$ bits/s/Hz. We can first observe that the required system transmit
power under different algorithms is reduced significantly as
the number of transmit antennas or reflecting units increases. Moreover,
under the same communication rate requirement, the proposed
algorithm leads to the lowest transmit power. The simulation results show
that the transmit power of the proposed algorithm is 20\% below that of the MRT-based scheme,
which demonstrates that introducing the IRS and the AN can offer significant
benefits.

\begin{figure}[t]
\centering{}\includegraphics[scale=0.55]{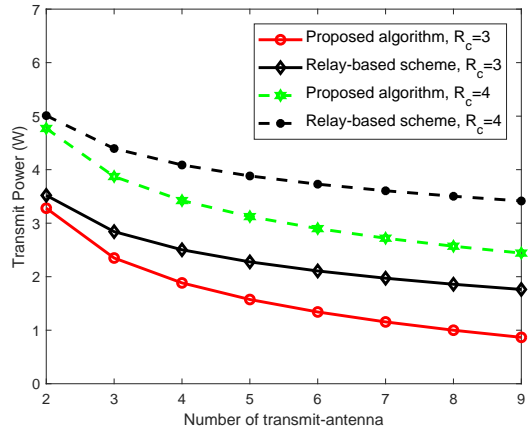}\caption{System transmit power versus the number of transmit antennas compared
with the relay-based scheme.}
\end{figure}

Fig. 7 compares the proposed algorithm with the relay-based scheme.
As the number of transmit antennas increases, the BS transmit power
required by the proposed algorithm is 25\% below that of the relay-based scheme. When the minimum communication rate requirement
$R_{c}$ increases, the transmit power in the proposed algorithm has increased but it is still lower than the relay-based scheme.

\begin{figure}
\centering{}\includegraphics[scale=0.34]{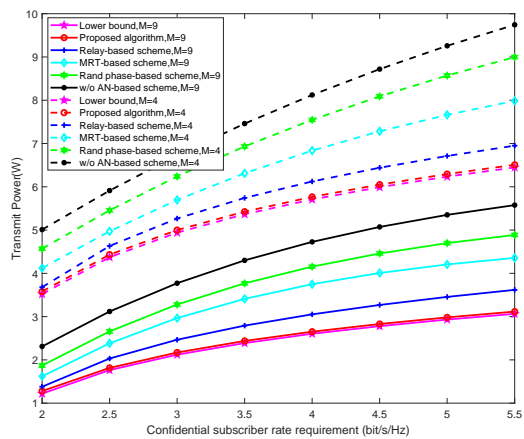}

\caption{System transmit power versus the confidential subscriber rate requirement (bit/s/Hz).}
\end{figure}

In Fig. 8, we investigate the impact of the confidential subscriber's minimum rate requirement on the BS transmit power under different
schemes. As the minimum rate requirement increases, the BS transmit
power corresponding to the different schemes also increases. The proposed algorithm outperforms all the baseline schemes in terms of power consumption. Under different numbers of transmit
antennas, all the schemes exhibit similar trends; however, as the confidential
subscriber rate requirement increases, the gap between the proposed
algorithm and the comparison schemes increases. In addition, we compare the proposed algorithm with its lower bound. The transmit power achieved by the proposed algorithm approaches the optimal power results, indicating the effectiveness of the proposed algorithm.

\begin{figure}
\centering{}\includegraphics[scale=0.55]{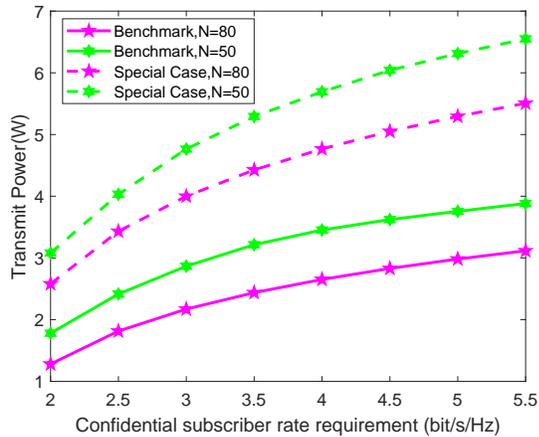}

\caption{System transmit power under different setups versus the confidential subscriber rate requirement (bit/s/Hz).}
\end{figure}
It is worth studying whether the system can realize a secure transmission when Alice is closer to the IRS. The confidential subscriber is located at $\left(50,0\right)$. We assume that Alice is fixed at $\left(30, 0\right)$, the IRS, Alice and the confidential subscriber are in the same link and Alice is close to the IRS. This special scenario is termed the ``Special Case''. We regard it as the benchmark that Alice is located at $\left(40,-30\right)$. As shown in Fig. 9, we investigate the impact of the confidential subscriber's minimum rate requirement on the BS transmit power in two different scenarios. Compared with the benchmark, the BS will consume more transmit power to achieve the same security rate of information transmission. However, it can still achieve the security transmission in the ``Special Case''. When Alice's position remains unchanged, increasing the number of reflection units at the IRS can effectively reduce the transmit power, which demonstrates the great potential of the IRS-aided scheme.

\section{Conclusion}

In this paper, we proposed an IRS-aided security classification wireless
communication solution to achieve secure and green wireless transmission.
Based on users' different security communication requirements,
we designed a security-selective transmission mechanism that reduces
unnecessary energy consumption. In the proposed scheme, the IRS serves
as an auxiliary link to assist communication users in secure information
transmission. The AN is mixed into the transmit signal, and then the
BS designs the beamforming vectors to perform the downlink transmissions.
To minimize the BS transmit power, we developed an alternating optimization
framework and employed SDR to transform the original nontrivial problem
into LMI problems. Then, we designed a SCA method to address the intractable
QoS constraints with respect to the beamformers. The simulation results
demonstrated the superiority of the proposed scheme over the baseline schemes in terms of transmit power consumption. These results show
the potential of the proposed scheme for achieving secure and
green information transmission.

\begin{appendices}

\section{Proof of Proposition 1}

If the optimal solution can be attained in each iteration, then we have
\begin{align}
 & \quad P_{t}^{\left(t\right)}\left(\mathbf{w}_{1}^{\left(t\right)},\mathbf{w}_{2}^{\left(t\right)},\mathbf{z}^{\left(t\right)},\boldsymbol{\Phi}^{\left(t\right)}\right)\nonumber \\
 & \geq P_{t}^{\left(t+1\right)}\left(\mathbf{w}_{1}^{\left(t+1\right)},\mathbf{w}_{2}^{\left(t\right)},\mathbf{z}^{\left(t\right)},\boldsymbol{\Phi}^{\left(t\right)}\right)\nonumber \\
 & \geq P_{t}^{\left(t+1\right)}\left(\mathbf{w}_{1}^{\left(t+1\right)},\mathbf{w}_{2}^{\left(t+1\right)},\mathbf{z}^{\left(t+1\right)},\boldsymbol{\Phi}^{\left(t\right)}\right)\nonumber \\
 & \geq P_{t}^{\left(t+1\right)}\left(\mathbf{w}_{1}^{\left(t+1\right)},\mathbf{w}_{2}^{\left(t+1\right)},\mathbf{z}^{\left(t+1\right)},\boldsymbol{\Phi}^{\left(t+1\right)}\right)\nonumber \\
 & =P_{t}^{\left(t\right)}\left(\mathbf{w}_{1}^{\left(t+1\right)},\mathbf{w}_{2}^{\left(t+1\right)},\mathbf{z}^{\left(t+1\right)},\boldsymbol{\Phi}^{\left(t+1\right)}\right),
\end{align}
where $t$ is the iteration index in Algorithm 1. Since the objective
$P_{t}^{\left(t\right)}$ is monotonically decreasing and lower bounded
due to the fundamental QoS requirements, the proposed algorithm is
guaranteed to converge.

\end{appendices}

\begin{IEEEbiography}[{\includegraphics[width=1in,height=1.25in,clip,keepaspectratio]{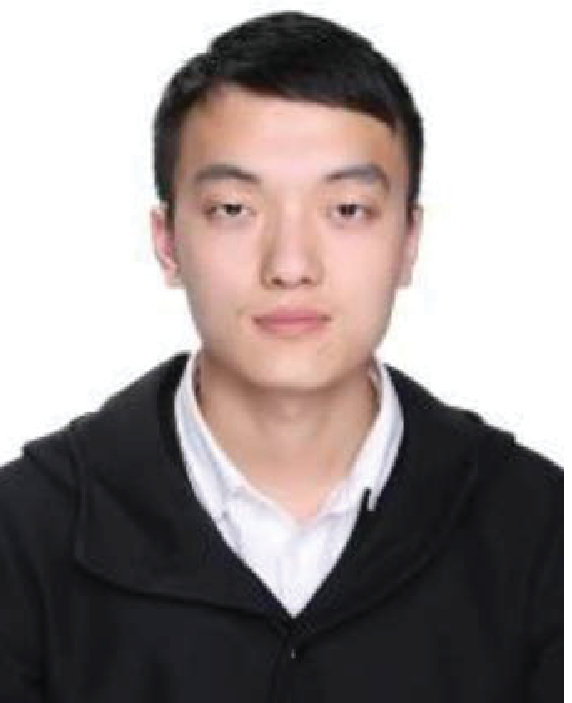}}]{Jintao Xing}
(S'18) received the B. S. degree from Wuhan University of Technology, Wuhan, China, in 2016. He is currently pursuing the Ph.D. degree in communication engineering with the School of Information and Communication Engineering, Beijing University of Posts and Telecommunications (BUPT), Beijing, China. His current research interests include physical layer security, wireless communication, intelligent reflecting surface and cell-free networks.
\end{IEEEbiography}

\begin{IEEEbiography}[{\includegraphics[width=1in,height=1.25in,clip,keepaspectratio]{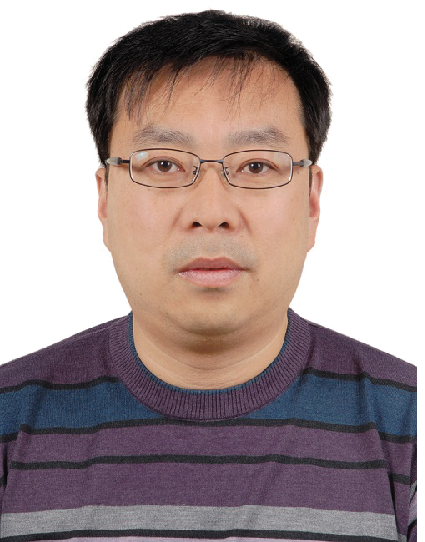}}]{Tiejun Lv}
(M'08-SM'12) received the M.S. and Ph.D. degrees in electronic engineering from the University of Electronic Science and Technology of China (UESTC), Chengdu, China, in 1997 and 2000, respectively. From January 2001 to January 2003, he was a Postdoctoral Fellow with Tsinghua University, Beijing, China. In 2005, he was promoted to a Full Professor with the School of Information and Communication Engineering, Beijing University of Posts and Telecommunications (BUPT). From September 2008 to March 2009, he was a Visiting Professor with the Department of Electrical Engineering, Stanford University, Stanford, CA, USA. He is the author of three books, more than 100 published IEEE journal papers and 200 conference papers on the physical layer of wireless mobile communications. His current research interests include signal processing, communications theory and networking. He was the recipient of the Program for New Century Excellent Talents in University Award from the Ministry of Education, China, in 2006. He received the Nature Science Award in the Ministry of Education of China for the hierarchical cooperative communication theory and technologies in 2015.
\end{IEEEbiography}

\begin{IEEEbiography}[{\includegraphics[width=1in,height=1.25in,clip,keepaspectratio]{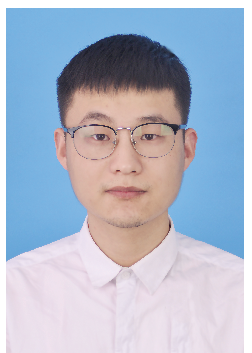}}]{Yashuai Cao}
(S'18) received the B.S. degree from Chongqing University of Posts and Telecommunications (CQUPT), Chongqing, China, in 2017. He is currently pursuing the Ph.D. degree in communication engineering with the School of Information and Communication Engineering, Beijing University of Posts and Telecommunications (BUPT), Beijing, China. His current research interests include wireless resource allocation and signal processing technologies for massive MIMO systems and intelligent reflecting surface assisted wireless networks.
\end{IEEEbiography}

\begin{IEEEbiography}[{\includegraphics[width=1in,height=1.25in,clip,keepaspectratio]{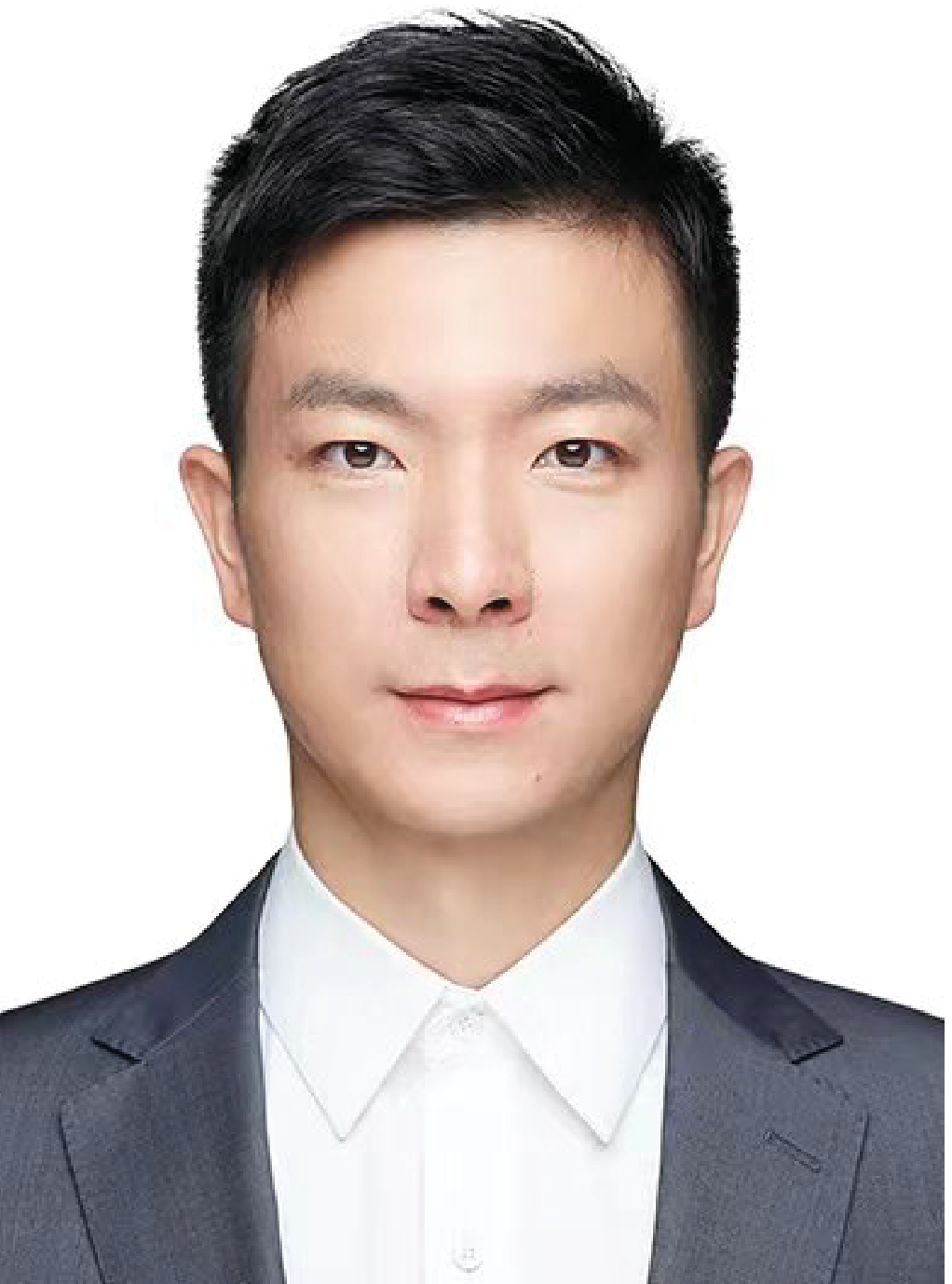}}]{Jie Zeng}
(M'09--SM'16) received the B.S. and M.S. degrees from Tsinghua University in 2006 and 2009, respectively, and received two Ph.D. degrees from Beijing University of Posts and Telecommunications in 2019 and the University of Technology Sydney in 2021, respectively.
From July 2009 to May 2020, he was with the Research Institute of Information Technology, Tsinghua University. From May 2020 to April 2022, he was a postdoctoral researcher with the Department of Electronic Engineering, Tsinghua University. Since May 2022, he has been an associate professor with the School of Cyberspace Science and Technology, Beijing Institute of Technology.
His research interests include 5G/6G, URLLC, satellite Internet, and novel network architecture. He has published over 100 journal and conference papers, and holds more than 40 Chinese and international patents. He participated in drafting one national standard and one communication industry standard in China.
He received Beijing's science and technology award of in 2015, the best cooperation award of Samsung Electronics in 2016, and Dolby Australia's best scientific paper award in 2020.
\end{IEEEbiography}

\begin{IEEEbiography}[{\includegraphics[width=1in,height=1.25in,clip,keepaspectratio]{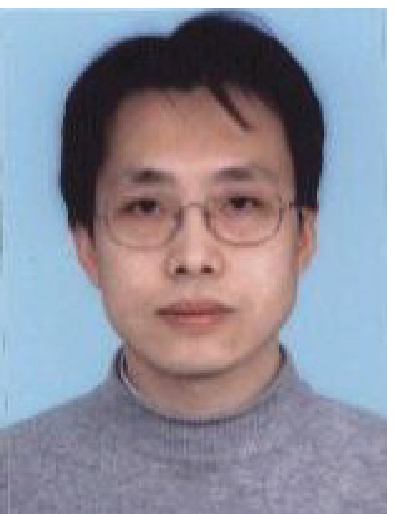}}]{Pingmu Huang}
Lecturer, School of Artificial Intelligence, Beijing University of Posts and Telecommunications. He received the M.S. degrees from Xi'an Jiaotong University, Xian, China, in 1996 and received Ph.D. degree of Signal and Information Processing from Beijing University of Posts and Telecommunications (BUPT), Beijing, China, in 2009. His current research interests include machine learning and signal processing. He published more than ten journal papers and conference papers on signal processing and machine learning.
\end{IEEEbiography}
\end{document}